\documentclass[journal, letterpaper, twoside, 10pt]{IEEEtran}

\usepackage{color}
\usepackage{amsmath, amssymb, amsthm} 
\usepackage{url}
\usepackage{cite}	

\newcommand*{\QEDA}{\hfill\ensuremath{\qedsymbol}}%

\ifCLASSINFOpdf
\else
  \usepackage[dvips]{graphicx}
  \graphicspath{{Figures/}}
\fi
\usepackage{epstopdf}
\begin{document}
%
\title{Packet Error Rate Analysis of Uncoded Schemes in Block-Fading Channels using Extreme Value Theory}
%

\author{Aamir~Mahmood~and~Riku~J{\"a}ntti,~\IEEEmembership{Senior~Member,~IEEE}%
\thanks{Manuscript received May 23, 2016; revised July 25 and September 15, 2016; accepted September 24, 2016.}%
\thanks{A.~Mahmood is with the Department
of Information and Communication Systems, Mid Sweden University, SE-851 70 Sundsvall, Sweden (e-mail: aamir.mahmood@miun.se).}%
\thanks{R.~J{\"a}ntti is with the Department of Communications and Networking, Aalto University School of Electrical Engineering,
Finland (e-mail: riku.jantti@aalto.fi).}
}

%
%

\markboth{IEEE Communications Letters}%
{Mahmood \MakeLowercase{\textit{et al.}}: PER Analysis of Uncoded Schemes in Block-Fading Channels using Extreme Value Theory}

%



\maketitle

\begin{abstract}
We present a generic approximation of the packet error rate (PER) 
function of uncoded schemes in the AWGN channel using 
extreme value theory (EVT). The PER function can assume both the exponential 
and the Gaussian $Q$-function bit error rate (BER) forms. The EVT 
approach leads us to a best closed-form approximation, in terms of 
accuracy and computational efficiency, of the average PER in block-fading 
channels. The numerical analysis shows that the approximation holds tight for 
any value of SNR and packet length whereas the earlier studies approximate the 
average PER only at asymptotic SNRs and packet lengths. 
\end{abstract}


%
\IEEEpeerreviewmaketitle


\section{Introduction}

\IEEEPARstart{I}{N} packet radio systems, the packet error rate (PER) analysis 
has practical significance for reliability and throughput estimation. Other 
than noise and packet collisions, therein, modeling the packet errors due to 
present fading conditions is an important problem. The time scale of fading 
relative to the bit or packet duration influences the selection of a packet error 
model. If the wireless channel remains constant across the length of a packet 
and the consecutive packets observe independent channel realizations, the packet 
error model must assume constant (block) fading for all the bits in a 
packet.     

Block-fading characterizes the wireless channels that experience slowly 
varying fading conditions. The PER evaluation in block-fading 
channels is studied extensively. However, for any 
modulation and coding scheme, its exact evaluation is complex and usually 
loose bounding methods as Jenson's inequality and Chernoff upper bound are 
employed~\cite{liu2012tput}.   

In \cite{xi2011general}, the average PER in 
Rayleigh block-fading, for both uncoded (commonly used in short-range 
radio systems) and coded schemes, is 
tightly upper bounded by $1-\mathrm{exp}(-\omega_0/\bar{\gamma})$, where $\bar{\gamma}$
 is average signal to noise ratio (SNR) and $\omega_0$ corresponds to the
inverse coding gain. By definition, $\omega_0 = \int_0^\infty \acute{P}_{e}(
\gamma)d\gamma$, where $\acute{P}_{e}(\gamma)$ is the PER in the AWGN channel. A 
log-domain linear approximation of $\omega_0$, for uncoded FSK, is 
proposed in \cite{liu2012tput}. In \cite{wu2014energy}, the same model is 
utilized to estimate the parameters of $\omega_0$ for different modulation 
schemes. However, the upper bound in 
\cite{xi2011general} and the approximations in 
\cite{liu2012tput}\cite{wu2014energy} are tight only in asymptotic regime: 
that is, when the average SNR is high or the packet length is large. 

In this paper, we present a new analytical method to approximate the PER function 
in the AWGN channel for uncoded schemes. The method uses the extreme value 
theory (EVT) to find a limiting distribution of the PER. The 
limiting distribution is easily integrable over Nakagami-$m$ 
fading distribution, and yields a best approximation of the average PER 
in block-fading channels. Our method offers an alternative approach to 
accurately approximate the average PER of uncoded schemes 
which either cannot be derived in closed-form or involve computationally 
extensive calculations.



\section{System Model}

Let $\mathbf{x}$ be the bits in a packet transmitted over a block-fading 
channel and $\mathbf{y}$ be the received symbols. Then, the input $\mathbf{x}$ and 
the output $\mathbf{y}$ of the channel are related as,
\begin{equation}
\mathbf{y} = h\mathbf{x}+\mathbf{n}
\label{eq:channelIO}
\end{equation}
where $h$ is the instantaneous fading coefficient. It is a zero-mean circularly 
symmetric complex Gaussian (CSCG) random variable with unit-variance 
i.e., $\mathbb{E}[\left|h\right|^2]=1$.
While, $\mathbf{n}$ is a sequence of mutually 
independent, zero-mean, CSCG noise with power $N_0$. 
If $P$ is the power per transmit bit, $\gamma = \left|h\right|^2 P/N_
0$ is the instantaneous received SNR and the average SNR is,
\begin{equation}
\bar{\gamma} = \mathbb{E}\left[\gamma\right] = \mathbb{E}\left[\left|h\right|^
2 P/N_0\right] = P/N_0 
\label{eq:avgSNR}
\end{equation}

The average PER over a block-fading channel, $\bar{P}_{e}(\bar{\gamma})$, is 
computed by integrating the PER in the AWGN, denoted as $\acute{P}_{e}(\gamma)$, 
over the distribution of received SNR, $p(\gamma; \bar{\gamma})$~\cite{xi2011general},    
\begin{equation}
	\bar{P}_{e}(\bar{\gamma}) = \int_{0}^{\infty}\acute{P}_{e}(\gamma)p(\gamma; \bar{\gamma})d\gamma.
\label{eq:p_avg}
\end{equation}
Depending on the radio propagation environment, the fading magnitude $|h|$ and consequently $\gamma
$ will take different distributions. Whereas, for an $N$-bit packet, $\acute{P}_{e}(\gamma
)$ is defined as,
\begin{equation}
\acute{P}_{e}(\gamma) = 1-\Big(1-b_e\left(\gamma\right)\Big)^{N}
\label{eq:p_awgn}
\end{equation}
where $b_e(\gamma)$ is the modulation-dependent instantaneous BER. 

We consider $b_
e(\gamma)$ with the following generic forms,
\begin{equation}
	b_{e}(\gamma) = c_mQ\left(\sqrt{k_m\gamma}\right) 
\label{eq:f_sq}
\end{equation}
\begin{equation}
	b_{e}(\gamma) = c_m\mathrm{exp}\left(-k_m\gamma\right)
\label{eq:f_se}
\end{equation} 
where $c_m$ and $k_m$ are the modulation-specific constants. The modulation 
schemes such as M-ASK, M-PAM, MSK, M-PSK and M-QAM have the BER form of \eqref{eq:f_sq} where $Q(
\cdot)$ is the Gaussian $Q$-function. Whereas, FSK and DPSK non-coherent 
modulations are described by the BER form in \eqref{eq:f_se}\cite{xi2011general}. 

\section{EVT based PER Function Approximation}
The evaluation of $\bar{P}_{e}(\bar{\gamma})$ is difficult because of $
\acute{P}_{e}(\gamma)$ expression involving a polynomial of degree $N$. In 
what follows, we provide a generic asymptotic approximation of 
\eqref{eq:p_awgn} using extreme value theory (EVT) for the BER expressions 
described in \eqref{eq:f_sq} and \eqref{eq:f_se} that allows to evaluate $\bar
{P}_{e}(\bar{\gamma})$ easily.    

\textit{Proposition 1:} The PER in the AWGN channel for the BER functions in 
\eqref{eq:f_sq} and \eqref{eq:f_se}, as the packet length $N \rightarrow \infty$, is 
asymptotically approximated by the Gumbel distribution function for the minimum, i.e.,
\begin{equation}
\acute{P}_{e}(\gamma) \approx 1 - \mathrm{exp}\left(-\mathrm{exp}\left(-
\frac{
\gamma-a_N}{b_N}\right)\right)
\label{eq:proposition1}
\end{equation}
where $a_N$ and $b_N > 0$ are the norming constants.

\textit{Proof:} For a packet of length $N$, the PER in \eqref{eq:p_awgn} for both 
\eqref{eq:f_sq} and \eqref{eq:f_se} can be written as,
\begin{equation}
\acute{P}_{e}(\gamma)  = 1 - \Big(1 + c_m\left(F_X\left(x\right)-1\right)\Big)
^{N} 
\label{eq:Prop1_trans}
\end{equation}
where $F_X(x)$ is the cumulative distribution function (CDF) of the standard normal distribution with $x=\sqrt{k_m
\gamma}$ for the BER in \eqref{eq:f_sq} while it is the CDF of the 
exponential distribution with $x=\gamma$ and $\lambda=k_m$ for 
\eqref{eq:f_se}. 

For $c_m =1$ (i.e., BPSK modulation), 
\eqref{eq:Prop1_trans} can be expressed in terms of $[F_X(x)]^{N}$ which can 
be approximated, as discussed later, with an extreme value 
distribution. In order to extend the same approach to 
any $c_m$, we manipulate \eqref{eq:Prop1_trans} as,
\begin{align}
\acute{P}_{e}(\gamma) & \approx 1 - \Big(1 + c_m\log F_X(x)\Big)^{N}  \nonumber \\
											& \approx 1 - \Big(e^{c_m \log F_X(x)}\Big)^{N} \nonumber \\
											& \approx 1 - \left[F_X(x)\right]^{Nc_m}
\label{eq:Prop1_trans2}
\end{align}
where $\log(\cdot)$ is the natural logarithm. Here, the first approximation 
results from the inequality $\log y \leq y -1$ for $y>0$, the second from $e^y 
\geq 1 + y$ and the last approximation uses the identity $y^a = e^{a\log y}$. The 
approximation 
in  \eqref{eq:Prop1_trans2} is quite tight and its 
accuracy increases as $N$ and $\gamma \rightarrow \infty$. 

By letting $N' = Nc
_m$, we handle $\left[F_X(x)\right
]^{N'}\!\!$ in \eqref{eq:Prop1_trans2} separately. Note that,
\begin{equation}
[F_X(x)]^{N'} \!\!\!\!\!\!\ =\prod_{i=1}^{N'}\Pr(X_i \leq x)=\Pr\left(
\underset{1 \leq i \leq {N'}}{
\max}\{X_i\}\leq x\right) 
\label{eq:max_evt}
\end{equation}
implies that $[F_X(x)]^{N'}\!\!\!$ can be approximated by one of the 
extreme value distributions: Gumbel, Fr{\'e}chet, Weibull 
\cite{song2006asymptotic}. To assimilate this, assume $X_1,\cdots,X_N$ be the 
i.i.d. random 
variables drawn from a common 
distribution function $F(x)$. Let $M_N = 
\overset{N}{\underset{i=1}{\max}}X_i$ denotes the maximum of first $N$ 
random variables. If there exist constants $a_N$ and $b_N > 0$, and a 
non-degenerate limit distribution $G(x)$ such that the CDF of normalized $M_n$
converges to $G(x)$,
\begin{equation}
{\underset{N\rightarrow \infty}{\lim}} \Pr\left(\frac{M_N - a_N}{b_N} \leq x
\right) {\rightarrow} G(x)
\label{eq:limit_dist}
\end{equation}
then $F(x)$ is said to be in the domain of attraction of $G(x)$. 

In order to find the exact limiting distribution, we recall the sufficient 
condition, stated in \cite{song2006asymptotic}\cite{evtBook}, for a distribution 
function $F(x)$ belonging to the domain of attraction of the Gumbel 
distribution. Let $\omega(F) = \mathrm{sup}\{x: F(x) < 1\}$, and assume that 
there is a real number $x_1$ such that for all $x_1\leq x < \omega(F), f(x) = 
F^{'}(x)$ and $F^{''}(x)$ exist, and $f(x) \neq 0$. If 
\begin{equation}
\underset{x\rightarrow\omega(F)}{\lim} \frac{d}{dx} \left[\frac{1-F(x)}{f(x)}
\right] = 0
\label{eq:evt_gumbel_condition}
\end{equation}  
then $F(x)$ converges to the Gumbel distribution as $N \rightarrow 
\infty$. The Gumbel distribution function is $G(x) = \mathrm{exp}(-\mathrm{exp}(
-\frac{\gamma-a_N}{b_N}))$ and constants $a _N$ and $b_N$ are,
\begin{equation}
\begin{aligned}
a_N & = F^{-1}\Big(1-\frac{1}{N}\Big) \\
b_N & = F^{-1}\Big(1-\frac{1}{Ne}\Big) - a_N
\end{aligned}
\label{eq:an_bn_basic}
\end{equation}
where $e$ is the base of the natural logarithm, and $F^{-1}(\cdot)$ denotes 
the inverse of $F(x)$. 

It is straightforward to show that the normal and the exponential distributions 
satisfy the sufficient condition stated in \eqref{eq:evt_gumbel_condition}. 
By replacing $[F_X(x)]^{N'}$ in \eqref{eq:Prop1_trans2} with the Gumbel 
distribution function completes the proof. \QEDA

The constants $a _N$ and $b_N$ for the normal distribution can be determined 
from \eqref{eq:an_bn_basic} by using $N = Nc_m$ and the quantile 
function of the normal distribution, $F^{
-1}\!(p)\!=\! \sqrt{2}~\mathrm{erf}^{-1}\!(2p-1)$. As $x = \sqrt{k_m\gamma}$, the transformation for $\gamma$ gives the constants,
\begin{equation}
\begin{aligned}
a_N & = \frac{2}{k_m} \Big(\mathrm{erf}^{-1}\Big(1 - \frac{2}{Nc_m}\Big)\Big)^
2  \\
b_N & = \frac{2}{k_m} \Big(\mathrm{erf}^{-1}\Big(1 - \frac{2}{Nc_me}\Big)\Big)
^2 - a_N
\end{aligned}
\end{equation}

Similarly for the exponential case, from \eqref{eq:an_bn_basic} and the quantile 
function of the exponential distribution, $F^{-1}(p;k_m)= -{\log (1-p)}/{k_m}$, we have,
\begin{equation}
a_N = \frac{\log (Nc_m)}{k_m}, b_N = \frac{1}{k_m}
\label{eq:exponential_aNbN}
\end{equation}

In next section, we show that the approximation in 
\eqref{eq:proposition1} can easily be evaluated under a general fading 
distribution.    

\section{PER over Block-Fading Channels}
We consider Nakagami-$m$ fading model for which the SNR, $\gamma$, is gamma distributed with the probability density function,
\begin{equation}
p(\gamma; \bar{\gamma}) = \frac{m^m \gamma^{m-1}}{\bar{\gamma}^m\Gamma(m)}\mathrm{exp}\Big(-
\frac{m\gamma}{\bar{\gamma}}\Big), \gamma \geq 0
\label{eq:Nakagami_Beta_Distr}
\end{equation}
where $0.5 \leq m <\infty$ is the fading parameter and 
$\Gamma\left(\cdot\right)$ is the standard gamma function \cite[p.892]{table2007}. 
Using \eqref{eq:proposition1} and \eqref{eq:Nakagami_Beta_Distr} in 
\eqref{eq:p_avg},
\begin{equation}
\bar{P}_e(\bar{\gamma}) \approx 1 - \frac{m^m}{{\bar{\gamma}}^m \Gamma(m)}\int_0^
\infty{e^{-e^{-\frac{\gamma - a_N}{b_N}}} \gamma^{m-1} e^{-\frac{m\gamma}{\bar
{\gamma}}}d\gamma}
\label{eq:PER_Integral_Naka}
\end{equation}

Let $s = \frac{m}{\bar{\gamma}}$ and $G(\gamma) = e^{-e^{-\frac{\gamma - a_N}{
b_N}}}$, then the integral in \eqref{eq:PER_Integral_Naka} is recognized as the Laplace transform of $\gamma^{
m-1}G\left(\gamma\right)$, i.e., 
\begin{equation}
\bar{P}_e(\bar{\gamma}) \approx 1 - \frac{m^m}{{\bar{\gamma}}^m \Gamma(m)} \int_0^
\infty{e^{-s\gamma} \gamma^{m-1} G(\gamma) d\gamma}.
\label{eq:PER_Integral_Naka_2}
\end{equation}
Using the Laplace transform of the PDF of the Gumbel distribution, $\mathcal{L}\{g(
\gamma)\} = e^{-a_Ns} \Gamma\left(1+ b_Ns\right)$, and the Laplace transform 
properties, $\mathcal{L}\{G(\gamma)\} = \frac{\mathcal{L}\{g(\gamma)\}}{s}$ and $\mathcal{L}\{t^n f(t)\} = (-1)^n \frac{d^n}{ds^n}\mathcal{L}\{f(t)\}$
, \eqref{eq:PER_Integral_Naka_2} is evaluated as,
\begin{equation}
\bar{P}_e(\bar{\gamma}) \!\approx \! 1 - \frac{m^m (-1)^{m-1}}{{\bar{\gamma}}^m \Gamma(m)} \frac{{d}^{m-1}}{{ds}^{m-1}}\!\! \left(\frac{e^{-s a_N} \Gamma\left(1+s b_N\right)}{s}\right)
\label{eq:FINAL_RESULT}
\end{equation}

For $m=1$ (Rayleigh fading), \eqref{eq:FINAL_RESULT} easily reduces to,
\begin{equation}
\bar{P}_e(\bar{\gamma}) \approx 1 - e^{-\frac{a_N}{\bar{\gamma}}} \Gamma\left(1+\frac{b_N}{\bar{\gamma}}\right).
\label{eq:rayA}
\end{equation}

By evaluating the derivative in \eqref{eq:FINAL_RESULT} for $m=2$ and $m=3$,
\begin{multline}
\bar{P}_e(\bar{\gamma}) \overset{m=2} \approx 1 - e^{-\frac{2a_N}{\bar{\gamma
}}} \Gamma\left
(\!1+\frac{2b_N}{\bar{\gamma}}\!\right) \! \bigg(\!1  + \frac{2a_N}{\bar{
\gamma}} - 
\frac{2b_N}{\bar{\gamma}} \times \\ \psi\left(\!1+\frac{2b_N}{\bar{\gamma}}
\!\right) \!\bigg)	
\label{eq:rayA_m2}
\end{multline}
\vspace{-15pt}
\begin{multline}
\bar{P}_e(\bar{\gamma}) \overset{m=3} \approx 1 - \frac{1}{2}e^{-\frac{3a_N}{\bar{\gamma
}}} \Gamma\!\left
(\!1+\frac{3b_N}{\bar{\gamma}}\!\right)\!\!
\Bigg(\!2 + \frac{9a_N^2}{\bar{\gamma}^2} + \frac{6a_N}{\bar{\gamma}} + \\
\qquad \qquad \,\,\, \frac{9b_N^2}{\bar{\gamma}^2}\psi\left(\!1, 1+\frac{3b_N}{
\bar
{\gamma}}\!\right)
- \frac{6b_N}{\bar{\gamma}}\psi\left(\!1+\frac{3b_N}{\bar{\gamma}}\!\right) \,
 +
 \\
  \qquad 	\qquad \frac{9b_N^2}{\bar{\gamma}^2}\psi\left(\!1+\frac{3b_N}{\bar{
\gamma}}
\!\right)^2 \!\!\!
- \frac{18a_Nb_N}{\bar{\gamma}}\psi\left(\!1+\frac{3b_N}{\bar{\gamma}}\!\right
)\!\!\Bigg)
\label{eq:rayA_m3}
\end{multline}
where $\psi(\cdot)$ and $\psi(1,\cdot)$ are the digamma and trigamma 
functions respectively.
 
\section{Numerical Results and Discussion}
We validate the proposed approximation of the average PER for uncoded FSK, BPSK and 16-QAM 
modulation schemes in block-fading. For this purpose, 
the approximated average PER in \eqref{eq:FINAL_RESULT} is validated against 
the numerical evaluation of \eqref{eq:p_avg}, and the approximation error is 
compared with earlier studies.  

\begin{figure}[!t]
	\vspace{-13pt}
\centering
	\includegraphics[width=1\linewidth]{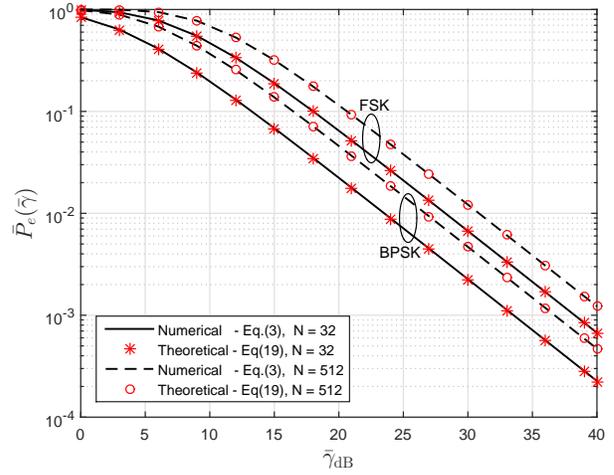}
	\vspace*{-1.7em}
    \caption{Average PER for FSK and BPSK modulations in Rayleigh 
		block-fading channel $\left(\mathrm{FSK}: c_m=1/2, k_m=1/2; \mathrm{BPSK}
: c_m=1, k_m=2 \right)$.}
	\label{fig:FSK_BPSK}
\end{figure}

\begin{figure}[!ht]
\vspace{-15pt}
\centering
	\includegraphics[width=1\linewidth]{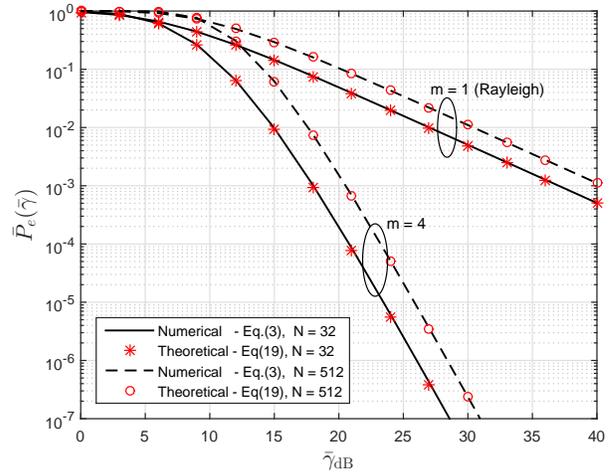}
    \vspace*{-1.7em}
		\caption{Average PER for uncoded 16-QAM in Nakagami-$m$ block-fading 
channel $\left(c_m = 3/4, k_m=4/5\right)$.}
	\label{fig:MQAM}
\end{figure}

The BER functions for non-coherent FSK and BPSK schemes are $\textstyle{\frac{1}{2}\mathrm{exp}
\left(-{\gamma}/2\right)}$ and $\textstyle{Q\left(\sqrt{2\gamma}\right)}$ respectively. 
The approximated average PER for these schemes is shown in 
Fig.~\ref{fig:FSK_BPSK} for $m = 1$, and depicts perfect matching to the numerical results 
for small and large packets. For 16-QAM, we use the BER 
approximation for M-QAM, $\tfrac{4}{k}\big(1-\tfrac{1}{\sqrt{M}}\big)Q\big(\scriptscriptstyle{\sqrt{
\tfrac{3k\gamma}{M-1}}}\big)$, where $M$ is the constellation size and $k \!=\! 
\log_2M$ is the number of bits per symbol \cite{simon2005digital}. The 
average PER for 16-QAM using \eqref{eq:FINAL_RESULT} shows the same accuracy over $m$, $N$ and $\bar{\gamma
}$ in Fig.~\ref{fig:MQAM}. 

It is worth noting that one can easily find an exact expression of the 
average PER for FSK by using the binomial expansion of \eqref{eq:p_awgn} with 
the BER in \eqref{eq:p_avg}. For instance, in Rayleigh fading with $p(\gamma; \bar
{\gamma}) = \mathrm{exp}(-\gamma/\bar{\gamma})$, we have,
\begin{equation}
\bar{P}_e\left({\bar{\gamma}}\right) = 1 - \sum_{n = 0}^{N} \binom{N}{n}
\left(-1
\right)^n \frac{(c_m)^n}{1 + nk_m\bar{\gamma}}.  
\label{eq:ExactPER_FSK}
\end{equation}
However, this expression requires the addition of terms, with alternating 
sign, each obtained from the multiplication of a very large number with a very 
small number. Consequently, it is numerically difficult to evaluate 
\eqref{eq:ExactPER_FSK} for large values of $N$. 

\begin{figure}[!ht]
\vspace{-15pt}
\centering
	\includegraphics[width=1\linewidth]{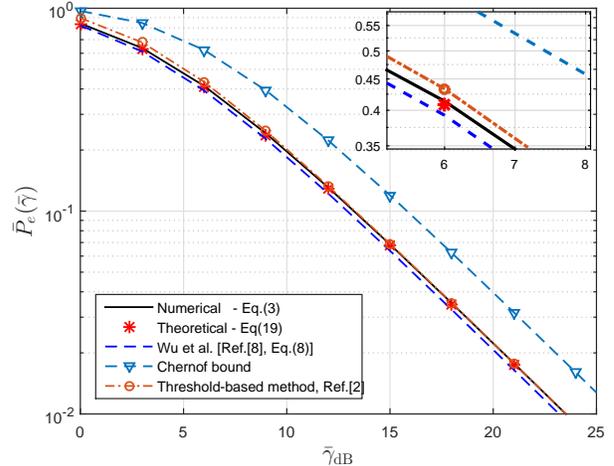}
	\vspace*{-1.7em}
    \caption{A comparison of different approximation techniques: 
average PER of BPSK modulation in Rayleigh block-fading with $N=32$.}
	\label{fig:BPSK_Low}
\end{figure}
Similarly, the Gaussian $Q$-function approximations (e.g.,~\cite{wu2011new,borjesson1979simple}) experience the same above-stated 
difficulty in evaluating the average PER of modulation schemes 
described by \eqref{eq:f_sq}. The other approach is to utilize the approximations 
for an integer power of the Gaussian $Q$-function (e.g., see \cite{Qn}). Such approximations are developed to evaluate the average 
symbol error probability (SEP) in fading with SEP in the form of $Q^N(x)$: e.g., differential encoded QPSK where maximum $N=4$. The approximation proposed in \cite{Qn} is integrable in Nakagami-$m$ fading for 
any $m$. However, it requires summation over 
all sequences of nonnegative integers $k_1,\cdots,k_{n_a}$ such that
$k_1\!+\cdots+k_{8}\! = \!N$, which is computationally demanding even for small 
values of $N$. In comparison, \eqref{eq:FINAL_RESULT} is easy to compute for any $N$.    

We analyze the average PER of BPSK in Rayleigh fading under the exponential 
function based approximations of the $Q$-function (Chernof bound and Wu \textit{et 
al.}~\cite{wu2011new} bound), the upper bound on PER 
\cite{xi2011general} and the proposed approximation in 
Fig.~\ref{fig:BPSK_Low}. Figure~\ref{fig:BPSK_Low} shows that the 
proposed approximation and upper bound \cite{xi2011general} match the 
numerical results tightly, and the latter bound is a good candidate 
for further comparison with \eqref{eq:FINAL_RESULT}.

For the BER in \eqref{eq:f_se}, the inverse coding gain or threshold, $w_0$, in the upper bound on 
average PER in Rayleigh fading, i.e., $\bar{P}_e({\bar{\gamma}}) \leq 1-\mathrm{exp}(-
\omega_0/\bar{\gamma})$ \cite{xi2011general}, is approximated in \cite{liu2012tput} as,
\begin{equation}
\omega_0 \approx k_1 \log N + k_2
\label{eq:Omega}
\end{equation}
where $k_1 \!=\! 1/k_m$ and $k_2 \!=\! \left(\gamma_e + \log c_m\right)/k_m$, and $\gamma_e = 0.5772$ is the Euler constant. In \cite{liu2012tput},
for BPSK, the Gaussian $Q$-function is approximated with an 
exponential function and $k_1$ and $k_2$ are calculated 
as above. In \cite{wu2014energy}, $\omega_0$ is estimated by fitting the 
linear model in \eqref{eq:Omega} to the simulations and, $k_1$ and $k_2$ are 
determined for different modulations. For 
uncoded 16-QAM, $k_1 \!=\! 2.327$ and $k_2 \!=\!-3.736$ 
(see \cite{wu2014energy}, Table~I).

\begin{figure}[!t]
\vspace{-14pt}
\centering
	\includegraphics[width=1\linewidth]{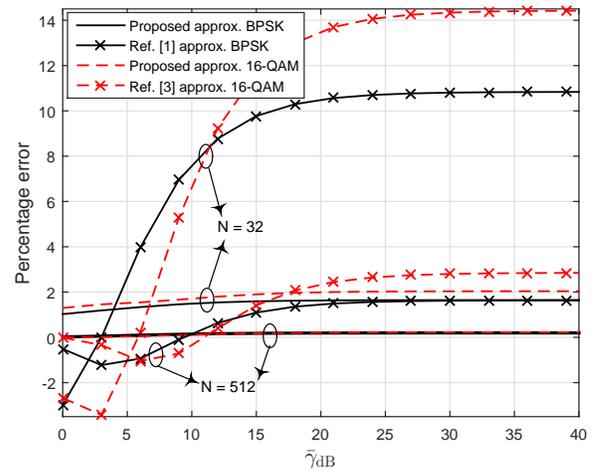}
	\vspace*{-1.9em}
    \caption{Approximation error in average PER for uncoded BPSK and 16-QAM 
schemes 
in Rayleigh block-fading channel.}
	\label{fig:comparison}
\end{figure}

In Fig.~\ref{fig:comparison}, the approximation error in \cite{liu2012tput}, \cite{wu2014energy} and  
\eqref{eq:FINAL_RESULT} for BPSK and 16-QAM in Rayleigh fading is compared.  
It can be observed that \eqref{eq:FINAL_RESULT} is quite accurate across 
any value of SNR and packet length. The error for 16-QAM is 
slightly higher than BPSK, because of the approximations in \eqref{eq:Prop1_trans2}, 
however this gap reduces quickly with increase in $N$. Figure~\ref{fig:comparison} also shows that 
\eqref{eq:FINAL_RESULT} is nearly insensitive to SNR and its accuracy 
mainly depends on the packet length. Whereas, 
both 
\cite{liu2012tput} and \cite{wu2014energy} yield large errors for small packets even at high SNRs. 
The errors in these studies decrease 
with increase in packet length but still remain higher than the proposed 
method. 
The approximation \eqref{eq:FINAL_RESULT} performs similarly for FSK, and the 
approximation error at $N=256$ is less than 0.04\%.   

\begin{figure}[!t]
\vspace{-13.5pt}
\centering
	\includegraphics[width=1\linewidth]{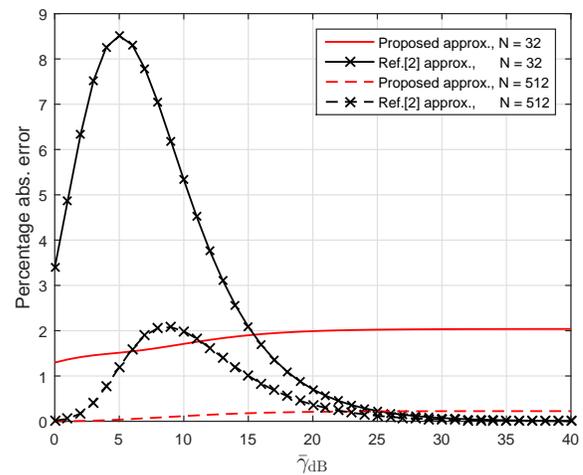}
	\vspace*{-1.9em}
    \caption{Approximation error in average PER for uncoded 16-QAM in Rayleigh 
		block-fading: comparing approximation \eqref{eq:FINAL_RESULT} 
with upper bound on PER \cite{xi2011general}.}
	\label{fig:comparisonO}
\end{figure}

In order to eliminate the effect of inaccuracies introduced by approximations of $w_0$ in 
\cite{liu2012tput} and \cite{wu2014energy}, we evaluate $w_0$ numerically and 
find the average PER from \cite{xi2011general}. For 16-QAM in 
Rayleigh fading, the resulting error is compared with the one in 
\eqref{eq:FINAL_RESULT} in Fig.~\ref{fig:comparisonO}. It can be seen that 
the upper bound in \cite{xi2011general} is accurate for large values of SNR and packet length while \eqref{eq:FINAL_RESULT} has much better accuracy at low SNRs for any 
packet lengths. Even at high SNRs, the error in \eqref{eq:FINAL_RESULT} is acceptably small and it reduces sharply with increase in packet length.
\section{Conclusions}
We showed that the PER of uncoded schemes, involving the 
BER forms of the exponential and the Gaussian $Q$-function, in the AWGN 
channel can be asymptotically approximated using EVT. Interestingly, the 
asymptotic distribution of the PER equals the Gumbel distribution function for 
sample minimum. For uncoded schemes, the EVT approach as compared to the 
threshold-based bound offers a  closed-form and more accurate
approximation of the average PER in block-fading channels, while latter approach
is still applicable for coded schemes.



%


\ifCLASSOPTIONcaptionsoff
  \newpage
\fi



%

\bibliographystyle{IEEEtran}
\bibliography{bib}

\end{document}